\documentclass[twocolumn,showpacs,preprintnumbers,amsmath,amssymb]{revtex4}

\usepackage{graphicx}% Include figure files
\usepackage{dcolumn}% Align table columns on decimal point
\usepackage{bm}% bold math

\newcommand{\bea}{\begin{eqnarray}} 
\newcommand{\beq}{\begin{equation}} 
\newcommand{\ear}{\end{array}} 
\newcommand{\eea}{\end{eqnarray}} 
\newcommand{\eeq}{\end{equation}}

\begin{document}

\title{\Large Superheavy Nuclei: Relativistic Mean Field Outlook.}

\author{A.\ V.\ Afanasjev \footnote{Email address: aafanasj@nd.edu}} 

\address{Department of Physics, University of Notre Dame, Notre Dame, 
Indiana 46556, USA}

\address{Department of Physics and Astronomy, Mississippi State
University, MS 39762, USA} 

\date{\today}

\begin{abstract}

  The analysis of quasiparticle spectra in heaviest $A\sim 250$ 
nuclei with spectroscopic data provides an additional constraint
for the choice of effective interaction for the description of
superheavy nuclei. It strongly suggest that only the parametrizations 
of the relativistic mean field Lagrangian which predict $Z=120$ and 
$N=172$ as shell closures are reliable for superheavy nuclei. The 
influence of the central depression in the density distribution of 
spherical superheavy nuclei on the shell structure is studied. 
Large central depression produces large shell gaps at $Z=120$ and
$N=172$. The shell gaps at $Z=126$ and $N=184$ are favored by a
flat density distribution in the central part of nucleus. It is
shown that approximate particle number projection (PNP) by means 
of the Lipkin-Nogami method removes pairing collapse seen at 
these gaps in the calculations without PNP.

\end{abstract}
 

\maketitle

  The possible existence of shell-stabilized superheavy nuclei, 
predicted with realistic nuclear potentials and the 
macroscopic-microscopic (MM) method (see references quoted in Ref.\ 
\cite{A250}), has been a driving force behind experimental and 
theoretical efforts to investigate superheavy nuclei. These investigations 
pose a number of challenges. The recent 
discovery of elements with $Z=115$ \cite{Z115} and $Z=116$ \cite{Z116} 
clearly shows great progress on the experimental side, but also indicates 
difficulties in the investigation of nuclei with low production cross 
sections and analyses based on few events.

  The theoretical challenges are also considerable since different
theoretical methods predict different spherical shell closures 
such as $Z=114, 120, 126$ for protons and $N=172, 184$ for neutrons 
\cite{RBM.02}. The largest variation in the predictions of shell closures 
appear in the self-consistent calculations based either on 
non-relativistic (Skyrme and Gogny \cite{BHR.03}) or relativistic 
(relativistic mean field [RMF] \cite{VRAL}) density functionals. 
Unfortunately, the properties of known superheavy nuclei in their 
ground states do not allow to discriminate between these predictions 
\cite{O.04}. This is due to the fact that known region of superheavy
elements is dominated by $\alpha$ decay, but $\alpha$-decays occur
between neighbouring nuclei and their half-lives are only insignificantly 
modified by shells effects \cite{GSI.04}.

  The part of these variations is definetely related to the fact that 
there is a large variety of the parametrizations for self-consistent 
models, but for many of them even the reliability of describing 
conventional nuclei is poorly known. In addition, self-consistent 
calculations have been confronted with experiment to a lesser degree
and for a smaller number of physical observables (mainly binding 
energies and quantities related to their derivatives) as compared 
with MM method. In particular, little attention has been paid to
single-particle degrees of freedom within the self-consistent models.  
However, the accuracy of predictions of spherical
shell closures depends sensitively on the accuracy of describing the
single-particle energies, which becomes especially important for 
superheavy nuclei where the level density is high. 

       Sect.\ \ref{A250-sect} shows how the study of quasiparticle 
states in odd-mass nuclei of the $A\sim 250$ region within the cranked 
relativistic Hartree-Bogoliubov (CRHB) theory \cite{CRHB} constrains 
the choice of the RMF parametrization for the study of superheavy
nuclei. The implications of such investigation for these 
nuclei are also discussed (Sect.\ \ref{A250-sect}). The influence of 
central depression in density
distribution on the shell structure of spherical superheavy nuclei
is discussed in Sect.\ \ref{Depression}. The importance of particle
number projection for the description of pairing properties of 
spherical superheavy nuclei is shown in Sect.\ \ref{Pairing}.
Finally, Sect.\ \ref{Concl} summarizes main conclusions.

%%%%%%%%%%%%%%%%%%%%%%%%%%%%%%%%%%%%%%%%%%%%%%%%%%%%%%%%%%%%%%%%
\section{The $A\sim 250$ mass region test \label{A250-sect}}
%%%%%%%%%%%%%%%%%%%%%%%%%%%%%%%%%%%%%%%%%%%%%%%%%%%%%%%%%%%%%%%%

   The investigation of the single-particle states in the odd-mass
deformed nuclei of the $A\sim 250$ mass region (the heaviest nuclei for
which detailed spectroscopic data are available) shed additional light 
on the reliability of the predictions of RMF theory on the energies of 
spherical subshells responsible for 'magic' numbers in superheavy nuclei. 
This is because several deformed single-particle states experimentally 
observed in odd nuclei of the $A\sim 250$ region originate from these 
subshells (see Table 3 in Ref.\ \cite{A250}).  

%%%%%%%%%%%%%%%%%%%%%%%%%%%%%%%%%%%%%%%%%%%%%%%%%%%%%%%%%%%%%%%%%%%%%%%%%%
\begin{figure}[!hbp]
\includegraphics[clip=true,scale=0.4]{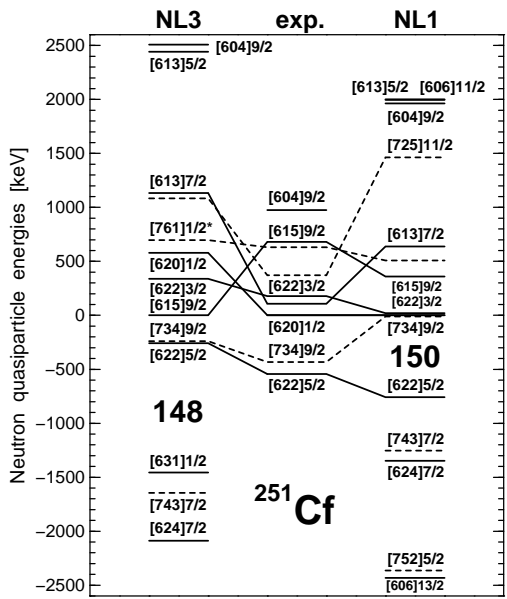}
%\bigskip
\caption{Experimental and theoretical quasiparticle energies of neutron
states in $^{251}$Cf. Positive and negative energies are used for particle
and hole states, respectively. Solid and dashed lines are used for positive 
and negative parity states, respectively. The symbols 'NL3' and 'NL1' 
indicate the RMF parametrization. From Ref.\ 
\cite{A250}}
\label{251Cf}
\end{figure}
%%%%%%%%%%%%%%%%%%%%%%%%%%%%%%%%%%%%%%%%%%%%%%%%%%%%%%%%%%%%%%%%%%%%%%%%%%

  The comparison between experimental and mean field single-particle
states is less ambiguous in deformed nuclei as compared with 
spherical ones \cite{MBBD.85,BM} at least at low excitation energies, 
where vibrational admixtures to the wave function are small. This is
a result  of the surface vibrations being less collective in deformed 
nuclei than in spherical ones since they are more fragmented 
\cite{Sol-book,MBBD.85}. As a consequence, the corrections to the 
energies of quasiparticle states in odd nuclei due to particle-vibration 
coupling are less state-dependent in deformed nuclei.

 A proper description of odd nuclei implies the loss of the time-reversal
symmetry of the mean-field, which is broken by the unpaired nucleon. The
BCS approximation has to be replaced by the Hartree-(Fock-)Bogoliubov 
method, with time-odd mean fields taken into account. The breaking of 
time-reversal symmetry leads to the loss of the double degeneracy 
(Kramer's degeneracy) of the quasiparticle states. This requires the use 
of the signature or simplex basis in numerical calculations, thus 
doubling the computing task. Furthermore, the breaking of the time-reversal 
symmetry leads to nucleonic currents, which cause {\it nuclear magnetism}
\cite{NM}. 
The CRHB(+LN) theory \cite{A190,CRHB} takes all these effects into 
account.

  First ever fully self-consistent description of quasiparticle states 
in the framework of the RMF theory was presented in Ref.\ \cite{A250} 
on the example of $^{249,251}$Cf and $^{249}$Bk nuclei. Fig.\ \ref{251Cf}
shows the CRHB results for $^{251}$Cf. Although the same set of quasiparticle 
states as in experiment appears, the calculated spectra are less dense. 
This is related to the effective mass (Lorentz mass in the notation of 
Ref.\ \cite{JM.89}) of the nucleons at the Fermi surface $m^*(k_F)/m$. 
While the experimental density of the quasiparticle levels corresponds 
to $m^*(k_F)/m$ close to one,  the low effective mass $m^*(k_F)/m \approx 
0.66$ of the RMF theory \cite{BRRMG.99} leads to a stretching of the 
energy scale. It has been demonstrated for spherical nuclei that the 
particle-vibration coupling brings the average level density in closer 
agreement with experiment \cite{MBBD.85}. Similar effect is expected
in deformed nuclei.

\begin{figure}[!hbp]
\includegraphics[clip=true,scale=0.5]{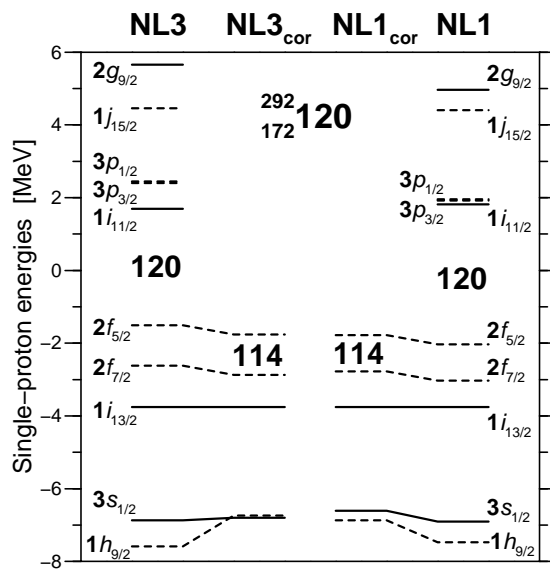}
%\bigskip
\caption{Proton single-particle states in a $^{292}_{172}120$ 
nucleus. Columns 'NL3' and 'NL1' show the states obtained in 
the RMF calculations at spherical shape with the indicated 
parametrizations. The energy of the $1i_{13/2}$ state in the 
NL1 parametrization is set to be equal to that in NL3, 
which means that the energies of all states in NL1 (last 
column) are increased by 0.78 MeV. The columns 'NL3$_{\rm cor}$' 
and 'NL1$_{\rm cor}$' show how the spectra are modified if 
empirical shifts were introduced based on discrepancies between 
calculations and experiment for quasiparticle spectra in 
deformed $^{249}$Bk. Solid and dashed lines are used for positive 
and negative parity states. Spherical gaps at $Z=114$ and 
$Z=120$ are indicated. From Ref.\ \cite{A250}}
\label{z120-sp}
\end{figure}

 The calculated energies of a number of states are rather close to experiment. 
On the other hand, the energies of some states and their relative positions 
deviate substantially from experiment. For example, only NL1 gives the 
correct ground state $\nu [620]1/2$ in $^{251}$Cf, whereas NL3
gives the $\nu [615]9/2$ (Fig.\ \ref{251Cf}). Detailed analysis shows 
that the discrepancies between experiment and calculations can be traced 
back to energies of spherical subshells from which deformed states emerge. 
This allows us to define 'empirical shifts' to the energies of spherical 
subshells (see Ref.\ \cite{A250} for details), which, 
if incorporated, will correct the discrepancies between calculations and 
experiment seen for deformed quasiparticle states. These `empirical shifts' 
are shown in Fig.\ \ref{z120-sp} as the energy difference between 
self-consistent and corrected energies of specific subshells.

%%%%%%%%%%%%%%%%%%%%%%%%%%%%%%%%%%%%%%%%%%%%%%%%%%%%%%%%%%%%%%%%%%%%%%%%%%
%%%%%%%%%%%%%%%%%%%%%%%%%%%%%%%%%%%%%%%%%%%%%%%%%%%%%%%%%%%%%%%%%%%%%%%%%%

  In the NL1 and NL3 parametrizations, the energies of the spherical
subshells, from which the deformed states in the vicinity of the Fermi
level of the $A\sim 250$ nuclei emerge, are described with an accuracy
better than 0.5 MeV for most of the subshells (see Fig.\ \ref{z120-sp}
in the present manuscript and Fig.\ 28 in Ref.\ \cite{A250}:
'empirical shifts', i.e. corrections, for single-particle energies
are indicated in both figures). The discrepancies (in the range of 
$0.6-1.0$ MeV) are larger for the $\pi 1h_{9/2}$ (NL3, NL1), 
$\nu 1i_{11/2}$ (NL3), $\nu 1j_{15/2}$ (NL1) and $\nu 2 g_{9/2}$ (NL3) 
spherical subshells. Considering that the RMF parametrizations were 
fitted only to bulk properties of spherical nuclei this level of 
agreement is good.
                                                                                
   In contrast, the accuracy of the description of single-particle 
states is unsatisfactory in the NLSH and NL-RA1 parametrizations, 
where 'empirical shifts' to the energies of some spherical subshells 
are much larger than in NL1 and NL3. NL-SH and NL-RA1 are the only 
RMF sets indicating $Z=114$ as a magic proton number \cite{LSRG.96,NL-RA1}. 
In the light of present results, these parametrizations should be 
considered as unreliable.

%%%%%%%%%%%%%%%%%%%%%%%%%%%%%%%%%%%%%%%%%%%%%%%%%%%%%%%%%%%%%%%%%%%%%%%%%%
\begin{figure}[!hbp]
\includegraphics[clip=true,scale=0.48]{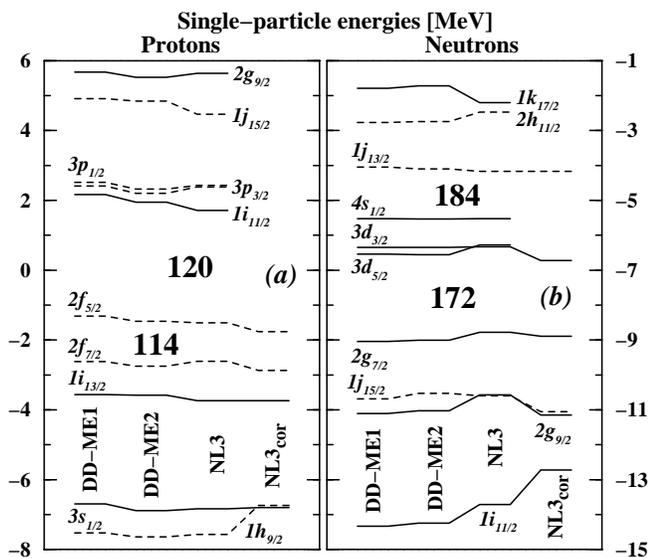}
%\bigskip
\caption{Single-particle spectra of the $^{292}$120
nucleus. Columns 'DD-ME2', 'DD-ME1' and 'NL3' show the states 
obtained in the RMF calculations at spherical shape with the 
indicated parametrizations. The column 'NL3$_{\rm cor}$' 
shows the spectra modified by empirical shifts. Solid and 
dashed lines are used for positive and negative parity states.}
\label{diff-forces}
\end{figure}
%%%%%%%%%%%%%%%%%%%%%%%%%%%%%%%%%%%%%%%%%%%%%%%%%%%%%%%%%%%%%%%%%%%%%%%%%%

   It is interesting to compare the results obtained through non-linear 
parametrizations of the RMF Lagrangian (NL1 and NL3) with the ones based 
on the parametrizations  which include an explicit density dependence of 
the meson-nucleon couplings (DD-ME1 \cite{DD-ME1} and DD-ME2 \cite{DD-ME2}). 
The latter parametrizations provide an improved description of asymmetric 
nuclear matter, nucleon matter, and nuclei far from stability \cite{DD-ME2}. 
However, the single-particle spectra of the $^{292}$120 nucleus obtained 
with DD-ME1 and DD-ME2 are similar to the ones seen in NL3 (Fig.\ 
\ref{diff-forces}). When compared with the NL3 spectra corrected for 
'empirical shifts' (columns NL3$_{cor}$ in Fig.\ \ref{diff-forces}, one 
can see that the DD-ME2 and DD-ME1 parametrizations do not remove the 
problems of the description of proton $\pi 1h_{9/2}$ state (Fig.\ 
\ref{diff-forces}a) and provide worse description of neutron $\nu 1i_{11/2}$ 
state as compared with NL3 set (see Ref.\ \cite{A250} for more detailed 
discussion of these states).
 
 The measured and calculated energies of the single-particle states at
normal deformation provide constraints on the spherical shell gaps of 
superheavy nuclei. Such analysis restricts the choice of the
RMF parametrizations only to those which predict $Z=120$ and $N=172$
as shell closures in superheavy nuclei \cite{A250}. In general, 
since the accuracy of the description of the $\nu 4s_{1/2}$ state is
unknown, we cannot also exclude the existence of the $N=184$ gap 
\cite{A250}. Similar analysis to Ref.\ \cite{A250} in non-relativistic 
models would restrict the choice of effective forces. 
However, it is already clear that the SkI4 Skyrme force which predicts 
$Z=114$ shell gap can be ruled out since it provides poor description 
of the spin-orbit splittings \cite{BRRMG.99}. Thus, one can conclude that 
non-relativistic theories suggest the existence of shell gaps (not 
necessary doubly shell closures) at $Z=120,126$ and $N=172,184$ 
(see Ref.\ \cite{AF.05} and references quoted therein). The role 
of self-consistency effects in the appearance of these shell gaps 
is discussed in the next section.

%%%%%%%%%%%%%%%%%%%%%%%%%%%%%%%%%%%%%%%%%%%%%%%%%%%%%%%%%%%%%%%
\section{Self-consistency effects in superheavy spherical 
nuclei \label{Depression}}
%%%%%%%%%%%%%%%%%%%%%%%%%%%%%%%%%%%%%%%%%%%%%%%%%%%%%%%%%%%%%%%

  Self-consistent microscopic calculations find a central depression 
in the nuclear density distribution \cite{BRRMG.99,DBDW.99}, which 
generates a wine-bottle nucleonic potential. The influence of this 
depression on the shell structure of spherical superheavy nuclei has 
been studied in Ref.\ \cite{AF.05} within the RMF theory without 
pairing. 

\begin{figure}[!hbp]
\includegraphics[clip=true,scale=0.3]{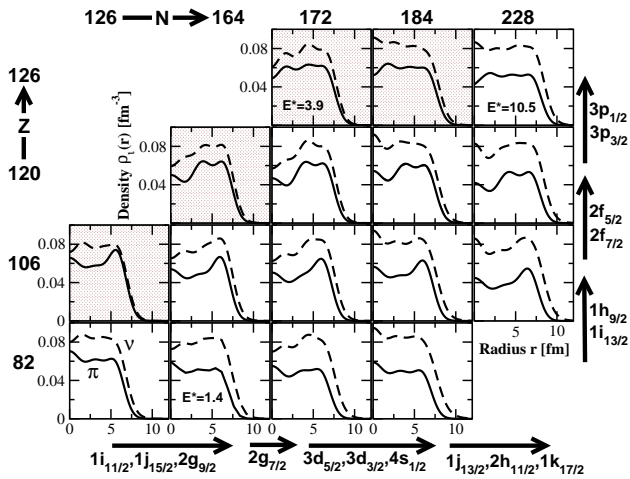}
%\bigskip
\caption{The evolution of proton and neutron densities
with the changes of proton and neutron numbers. Arrows indicate the group
of single-particle subshells which become occupied with the change of the 
nucleon number. The figure is based on the results of spherical RMF 
calculations without pairing employing the NL3 \cite{NL3} parametrization. 
The shaded 
background is used for nuclei located beyond the proton-drip line. If the 
indicated configuration is not lowest in energy, its excitation energy (in 
MeV) is given by E*. From Ref.\ \protect\cite{AF.05}}
\label{dens-sys}
\end{figure}

  The underlying microscopic mechanism for an appearance of this central 
density depression is illustrated in Fig.\ \ref{dens-sys}. The starting 
point of this consideration is the density distribution in $^{208}$Pb,
which is nearly flat in central region of nucleus. Its charge 
distribution is well described by the RMF theory \cite{GRT.90,NL1}.
It is general feature of all nuclear structure models that on going 
from $^{208}$Pb to spherical superheavy nuclei in the region 
around $Z=120$ and $N=172$, the ground state configurations are built
first by the occupation of the group of high-$j$ subshells (neutron 
$\nu 1i_{11/2}$, $\nu 1 j_{15/2}$, and $\nu 2g_{9/2}$ and proton
$\pi 1i_{13/2}$ and $\pi 1 h_{9/2}$) and then by the occupation
of the group of medium-$j$ subshells (neutron $\nu 2g_{9/2}$ and
proton $2f_{7/2}$ and $2f_{5/2}$), see Fig.\ \ref{dens-sys}. The 
high-$j$ subshells are localized mostly near the surface, whereas 
the low-$j$ subshells have a more central localization. As a 
consequence of this grouping of high-$j$ and medium-$j$ subshells 
above $^{208}$Pb, the density is added mostly in the surface region 
which leads to the appearance of the central density depression in 
the nuclei around the $Z=120, N=172$ system, see Fig.\ \ref{dens-sys}.

%%%%%%%%%%%%%%%%%%%%%%%%%%%%%%%%%%%%%%%%%%%%%%%%%%%%%%%%%%%%%%%%%%%%%%%%
%%%%%%%%%%%%%%%%%%%%%%%%%%%%%%%%%%%%%%%%%%%%%%%%%%%%%%%%%%%%%%%%%%%%%%%%

  On the contrary, the group of low-$j$ subshells (neutron $\nu 3d_{5/2}$, 
$\nu 3d_{3/2}$ and $\nu 4s_{1/2}$ and proton $\pi 3p_{3/2}$ and 
$\pi 3p_{1/2}$) is filled on going from the $Z=120, N=172$ system to the 
$Z=126, N=184$ system. Since filling up a low$-j$ group with nucleons
increases the density near the center, the density distribution
in the central part of nucleus is nearly flat in the latter system. 

\begin{figure}[!hbp]
\includegraphics[clip=true,scale=0.4]{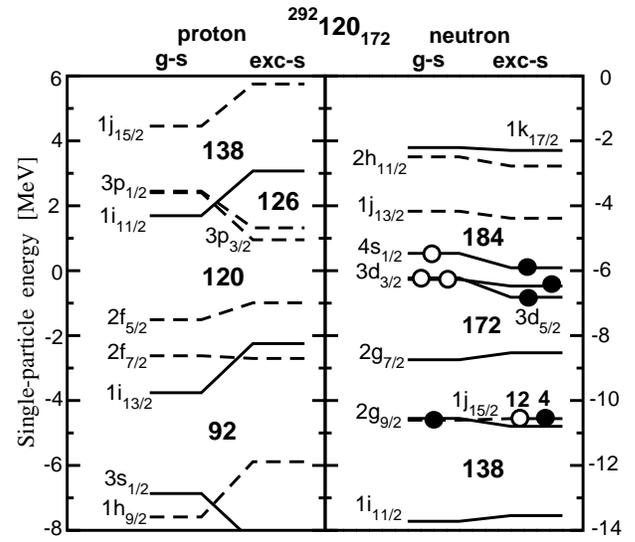}
%\bigskip
\caption{Single-particle spectra of the ground state (indicated 
as 'g-s') and the excited (indicated as 'exc-s') configurations
in the $^{292}120_{172}$ system obtained in the RMF calculations 
with the NL3 force. Solid and dashed lines are used for positive 
and negative parity, respectively. Solid and open circles indicate 
the occupied and empty subshells, respectively. In the ground state, 
all subshells below $Z=120$ and $N=184$ are fully occupied. In 
the excited configuration, only 12 particles are excited from
the subshell $\nu 1j_{15/2}$: 4 particles still reside in this
subshell. The spherical shell gaps of interest are indicated.
From Ref.\ \cite{AF.05}}
\label{spectra}
\end{figure}

  The magic gaps of the wine-bottle (the case of central depression 
in density distribution) and flat-bottom (the case of flat density
distribution) nucleonic potentials are different. This is illustrated
in Fig.\ \ref{spectra}, where starting from the ground state configuration
of $^{292}$120$_{172}$ nucleus, having a central depression (configuration 
'g-s'), a flatter density distribution in the central part of nucleus 
is generated by exciting particles from high-$j$ subshells to low-$j$ 
subshells (configuration 'exc-s') \cite{AF.05}. While wine-bottle potential 
(conf. 'g-s') is characterized by the large $Z=120$ and $N=172$ large gaps, 
we see the appearance of the $Z=126$ proton gap and the shrinking of 
the $Z=120$ shell gap in the flat-bottom potential (conf. 'exc-s'). 
To a lesser extend, the $N=172$ neutron gap decreases and the $N=184$ 
gap increases in the latter potential.

%%%%%%%%%%%%%%%%%%%%%%%%%%%%%%%%%%%%%%%%%%%%%%%%%%%%%%%%%%%%%%%%
%%%%%%%%%%%%%%%%%%%%%%%%%%%%%%%%%%%%%%%%%%%%%%%%%%%%%%%%%%%%%%%%%

   Due to the isovector force, which tries to keep the neutron and
proton density profiles alike, there  is a mutual enhancement of the
$Z=120$ and $N=172$ gaps, both being favored by the wine-bottle potential,
and of the $Z=126$ and $N=184$ gaps, both favored by the flat bottom
potential. For the same reason the gaps are smaller for the combination
$Z=126$ and $N=172$, and the $Z=120$ gap does not develop for the
$N=184$ systems. This behavior is not expected to depend much on the 
density functional chosen (see discussion in Ref.\ \cite{AF.05}). 

    Considering that $Z=120, 126$ and $N=172, 184$ are the particle
numbers which appear as the candidates for 'magic' particle numbers
in self-consistent theories, it is clear that the magnitude of the
central depression in density distribution is an important factor
in defining 'magic' shell gaps in spherical superheavy nuclei. 
This magnitude is correlated with effective mass $m^*/m$ of microscopic
theory: more pronounced central depression develops for the density
functionals with low effective mass \cite{BRRMG.99}. This feature 
may be understood as follows \cite{AF.05}. In the surface region, 
$m^*/m$ changes from low value ($<1$) in the interior to 1 in the 
exterior. Classically, nucleons with given kinetic energy are more 
likely to be found in regions with high effective mass than in the 
regions with low one because they travel with lower speed. This is 
reflected by the Thomas-Fermi expression for the nucleonic density 
$\rho \propto [2 m^*(\epsilon_F-V)]^{3/2}$. The increase of the 
effective mass in the surface region favors the transfer of mass 
from the center there, which makes the above discussed polarization 
mechanism of the high-$j$ subshells more effective for functionals 
with low effective mass.

  All experimentally known nuclei with $Z\geq 100$ are expected to 
be deformed \cite{HM.00,O.01}. The deformation leads to a more equal 
distribution of the single-particle states emerging from the high-$j$ 
and low-$j$ spherical subshells (see, for example, the Nilsson 
diagrams in Figs.\ 3-4 in Ref.\ \cite{CAFE.77}) than for spherical 
shape. Thus, the density profile of a deformed nucleus is relatively
flat \cite{AF.05,PXS.05}, strongly resembling the one of 
phenomenological potentials. In addition, the density profile variations 
as a function of particle number are less drastic than in spherical 
nuclei. These features together with the fact that the single-particle 
energies of the deformed nuclei in heavy actinide region have been 
carefully fitted in the phenomenological potentials explains the 
success of the shell correction method \cite{MN.94,PS.91} in the
description of known superheavy nuclei. However, this method neglects 
the self-consistent rearrangement of single-particle levels due to 
the appearance of a central depression  in spherical superheavy 
nuclei. Thus the predictions of the magic numbers for superheavy 
nuclei within the shell correction method should be  considered 
with caution.

%  THE UNCERTAINTIES IN EFFECTIVE MASS

%  NAZAREWICS RESULTS FOR SURFACE SKIN 

%%%%%%%%%%%%%%%%%%%%%%%%%%%%%%%%%%%%%%%%%%%%%%%%%%%%%%%%%
\section{Pairing correlations in superheavy nuclei. \label{Pairing}}
%%%%%%%%%%%%%%%%%%%%%%%%%%%%%%%%%%%%%%%%%%%%%%%%%%%%%%%%%

  Almost all published self-consistent calculations for superheavy 
nuclei performed either in an approximate HF+BCS or full 
Hartree-Fock-Bogoliubov (HFB) frameworks show the collapse of 
pairing at large shell gaps in spherical superheavy nuclei (see, 
for example, Refs.\ \cite{SPSCV.04,ZMZGT.05,LSRG.96}. 
Principal shortcoming of these calculations is the fact that 
neither the BCS nor the HFB wave functions are the eigenstates 
of particle number operator \cite{Ring-book,A250}.
The best way to deal with this problem would be to perform an 
exact particle number projection before the variation 
\cite{Ring-book}, but this is very time-consuming for realistic 
interactions. In this section, the importance of particle number 
projection on the properties of spherical superheavy nuclei
is studied employing the CRHB theory \cite{CRHB} with and without 
approximate particle number projection by means of the 
Lipkin-Nogami (LN) method.

 The CRHB(+LN) calculations are performed using the NL3 
parametrization for the RMF Lagrangian and Gogny D1S force
in the pairing channel. The scaling factors $f$ of the 
Gogny D1S force are selected as follows: $f=1.0$ in the 
calculations without LN (CRHB) and $f=0.864$ in the 
calculations with LN (CRHB+LN). These scaling factors 
provide good description of the moments of inertia of 
rotational bands in the $A\sim 250$ mass region 
\cite{A250}.
 
\begin{figure}[!hbp]
\includegraphics[clip=true,scale=0.46]{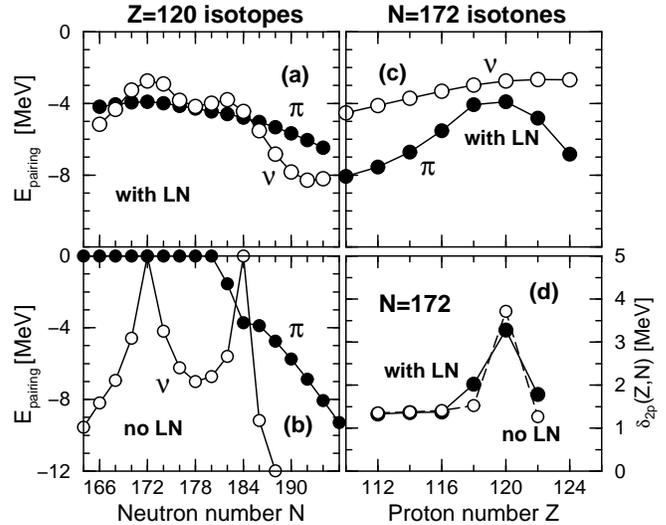}
%\bigskip
\caption{Pairing energies $E_{pairing}$ as a function 
of particle number obtained in the CRHB calculations with 
(panels (a) and (c)) and without (panel (b)) approximate 
particle number projection by means of the Lipkin-Nogami 
method for the $Z=120$ isotopes (panels (a) and (b)) and 
$N=172$ isotones (panel (c)). Panel (d) shows the 
$\delta_{2n}(Z,N)$ quantity for the $N=172$ isotones
obtained in the CRHB calculations with and without LN.}
\label{pairing}
\end{figure}

 Figs.\ \ref{pairing}a,b compare the calculated pairing
energies $E_{pairing}=-\frac{1}{2} Tr (\Delta \kappa)$
for $Z=120$ isotopes obtained in the CRHB and CRHB+LN 
calculations. The $Z=120$ gap is large in the vicinity
of $N=172$ (see Sect.\ \ref{Depression}) which leads to the
collapse of proton pairing in the CRHB calculations
(Fig.\ \ref{pairing}b).
With increasing neutron 
number, the densities in the central part of nuclei 
become flatter (see Sect.\ \ref{Depression}) leading to 
shrinking of the $Z=120$ gap. Because of that the
proton pairing shows up at $N=182$ and  increases
in absolute magnitude with the increase of $N$
(Fig.\ \ref{pairing}b).
Because neutron $N=172$ and $N=184$ gaps are smaller than the 
$Z=120$ gap (Fig.\ \ref{diff-forces}), the pairing 
collapse in neutron subsystem 
is seen only at these neutron numbers 
(Fig.\ \ref{pairing}b). Similar collapse 
of pairing is seen at 'magic' shell closures in 
the CRHB calculations for the $N=172$ and $N=184$ 
isotones (not shown here). On the contrary, no 
pairing collapse is observed in the CRHB+LN 
calculations for these chains of nuclei (Figs. 
\ref{pairing}a,c). In addition, smaller variations
of $E_{pairing}$ as a function of particle number
are seen in the CRHB+LN calculations as compared
with the CRHB ones.

   It was suggested in Ref.\ \cite{ZMZGT.05} to use the
fact that the pairing energies vanish at closed shells 
as a fingerprint of the shell gaps. Present studies in 
the CRHB+LN framework do not support this suggestion. 
One should also note that the LN method maybe less reliable 
in the regime of weak pairing \cite{AER.02} typical for large
shell gaps, so more detailed investigation of the pairing 
in spherical superheavy nuclei in the formalism with exact 
particle number projection is needed.

%%%%%%%%%%%%%%%%%%%%%%%%%%%%%%%%%%%%%%%%%%%%%%%%%%%%%%%%%%%%%%%
%%%%%%%%%%%%%%%%%%%%%%%%%%%%%%%%%%%%%%%%%%%%%%%%%%%%%%%%%%%%%%%%

    One of the observables used frequently in the search of 
shell gaps in superheavy nuclei is the $\delta_{2n,(2p)}(N,Z)$ 
quantity (called sometimes as two-nucleon shell gap [see
Ref.\ \cite{A250} for the discussion of this quantity]) 
which for the neutrons (and similarly for the protons) is 
defined as
\begin{eqnarray}
\delta_{2n}(Z,N)=S_{2n}(Z,N)-S_{2n}(Z,N+2) \nonumber
\label{2n-shell-gap}
\end{eqnarray}
This quantity is related to the derivative of the separation 
energy $S_{2n}(Z,N)$, and thus, it is a sensitive indicator of 
the localization of the shell gaps. As follows from the 
comparison of the CRHB and CRHB+LN calculations (Figs.\ 
\ref{pairing}c and d), the presence of pairing at shells 
gaps affects the $\delta_{2n,(2p)}(N,Z)$ quantity
in the following way: the peak in $\delta_{2n,(2p)}(N,Z)$
becomes more smooth and broad and the magnitude of it 
decreases.

\section{Conclusion  \label{Concl} }

   The analysis of quasiparticle spectra in 
heaviest $A\sim 250$ nuclei provides an additional constraint 
for the choice of effective interaction for the description
of superheavy nuclei within specific model.
Based on this analysis it was concluded
that only parametrizations of the RMF Lagrangian predicting
$Z=120$ and $N=184$ as shell closures provide reasonable 
description of the spectra of odd-mass nuclei. No support 
for the $Z=114$ shell gap has been established. One can 
restrict the predictions of self-consistent models (including 
non-relativistic ones) to shell gaps at $Z=120, 126$ and 
$N=172, 184$. The investigation of self-consistency effects 
related to central depression in the density distribution finds 
important correlations between these gaps: large central depression 
produces large shell gaps at $Z=120$ and $N=172$, while the shell 
gaps at $Z=126$ and $N=184$ are favored by a flat density 
distribution in the central part of nucleus. However, the magnitude 
of the central depression correlates with an effective mass of
nucleon in 
specific model/parametrization. Unfortunately, there is no consensus 
on what value an effective mass should have in self-consistent 
theories (see, for example, the discussion in Refs.\ \cite{FPT.01}).
Because of existing differences in effective mass of nucleon in 
different models/parametrizations (especially pronounced for 
Skyrme forces), the studies of quasiparticle spectra in deformed 
actinide nuclei may not provide sufficient constraint for localization 
of magic double shell closure (if any) in spherical superheavy nuclei. 
Approximate particle number projection (PNP) by means of the 
Lipkin-Nogami method removes pairing collapse seen at large shell 
gaps in spherical superheavy nuclei in the calculations without PNP. 
Since closed shell nuclei maybe in the regime of weak pairing, more 
detailed studies of the pairing properties of superheavy nuclei in 
the formalism with an exact particle number projection are needed.

\bigskip\noindent
{\large \bf Acknowledgements}

  This work was supported by the DOE grant DE-F05-96ER-40983. I would 
like to express my gratitude to S.\ Frauendorf, T.\ L.\ Khoo, 
G.\ Lalazissis, and I.\ Ahmad for their contributions into this
project.

\end{document}